 \documentclass[final,5p,times,twocolumn]{elsarticle}

\usepackage{amssymb}
\usepackage{amsmath}
\usepackage{float}
\biboptions{square, sort&compress}

\journal{International Journal of Mass Spectrometry}

\begin{document}

\begin{frontmatter}



\title{Wide-band mass measurements with a multi-reflection time-of-flight mass spectrograph}


\author[RIKEN,NMSU]{P. Schury}
\author[RIKEN]{Y. Ito}
\author[RIKEN]{M. Wada} 
\author[NMSU]{H. Wollnik}

\address[RIKEN]{RIKEN Nishina Center for Accelerator Physics, Wako, Japan}
\address[NMSU]{New Mexico State University, Department of Chemistry and BioChemistry, Las Cruces, NM, USA}

\begin{abstract}
We characterize the mass bandwidth of the multi-reflection time-of-flight mass spectrograph, showing both the theoretical and effective mass bandwidth.  We then demonstrate the use of a multi-reflection time-of-flight mass spectrograph to perform mass measurements in mass bands much wider than the mass bandwidth.

\end{abstract}


\end{frontmatter}

% \linenumbers

\section{Introduction}
\label{secIntro}

\par The multi-reflection time-of-flight mass spectrograph (MRTOF-MS), first proposed more than 20 years ago \cite{WollnikMRTOF}, uses a pair of electrostatic mirrors to compress a flight path of several hundred meters (or even many kilometers in some cases) within a reflection chamber of $\approx$1~m length.  With properly designed ion traps to quickly prepare ions, the MRTOF-MS can achieve mass resolving powers of $R_{\mathrm{m}}$$>$10$^5$ while operating at rates of 100~Hz or more \cite{SchuryEMIS}\cite{ItoEMIS}\cite{Plass}.  
\par In recent years, these devices have begun to prove useful for online measurement of nuclear masses\cite{KreimCaNature}\cite{ItoLi8}.  The technique has been demonstrated to be able to accurately provide mass precision on the level of $\delta m$/$m$ $\sim$ 5$\times$10$^{-7}$ or better.  With it's ability to achieve very high mass resolving powers while operating with very low intensities, the MRTOF-MS could become a useful instrument for analytical chemistry.
\par However, the multi reflection nature of the measurement has, thus far, made analysis of rich, wide-band mass spectra difficult or impossible.  Starting with a detailed exploration of the mass bandwidth of the MRTOF-MS, we have developed a technique that allows simple analysis of even wide-band mass spectra using the MRTOF-MS.  Employing such an analytic method, we believe the device could eventually provide wide-band measurements of nuclear masses much in the way of storage rings \cite{ESR}.  The device could similarly be useful in analytic chemistry, providing wide-band analysis much like FT-ICR Penning traps, but with a much greater sensitivity.

\section{MRTOF-MS Technique}
\label{secMRTOF}

\par The MRTOF-MS begins with an ion trap to prepare ions as well-cooled pulses \cite{ItoEMIS}.  Ion pulses extracted from the trap are then transferred to the reflection chamber.  The reflection chamber consists of a pair of electrostatic mirrors, a lens, and a field-free drift region.  The outmost electrode of each mirror is switchable, allowing ions to enter and leave the reflection chamber.   A multichannel plate (MCP) ion detector is mounted after the reflection chamber.

\begin{figure}[H]
 \includegraphics[width=0.48\textwidth]{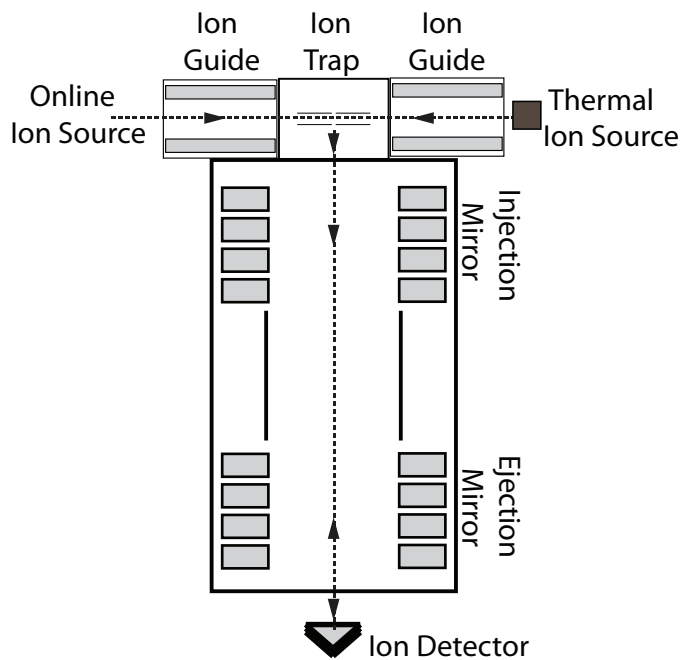}
% \vspace{-8mm}
\caption{Sketch of the MRTOF with ion source and preparation trap.  Not to scale.}
\label{figSketch}
\end{figure}

\par The signal to extract ions from the trap also serves as the start signal for a time-to-digital converter (TDC).  The potential on the outermost electrode of the injection-side mirror is reduced by $\approx$1~kV a few microseconds prior to issuing the signal to extract ions from the trap.  Ions then enter the reflection chamber, travel to the ejection-side mirror turning-point and return towards the injection-side mirror.  The potential on the outermost electrode of the injection-side mirror is returned to its nominal value before returning ions come close enough to sample the changing electric field.  Ions will then reflect between the mirrors until the potential of the outermost electrode of the ejection-side mirror is reduced.  Ions then will pass out of the reflection chamber and travel to the MCP, the signal from which serve as stop signals for the TDC.  The time of flight, given by the time between start and stop signals, is recorded and the cycle is repeated until sufficient statistics have been accumulated.

\par
\begin{figure}[h]
%\resizebox{0.54\textwidth}{!}{%
 \includegraphics[width=0.48\textwidth]{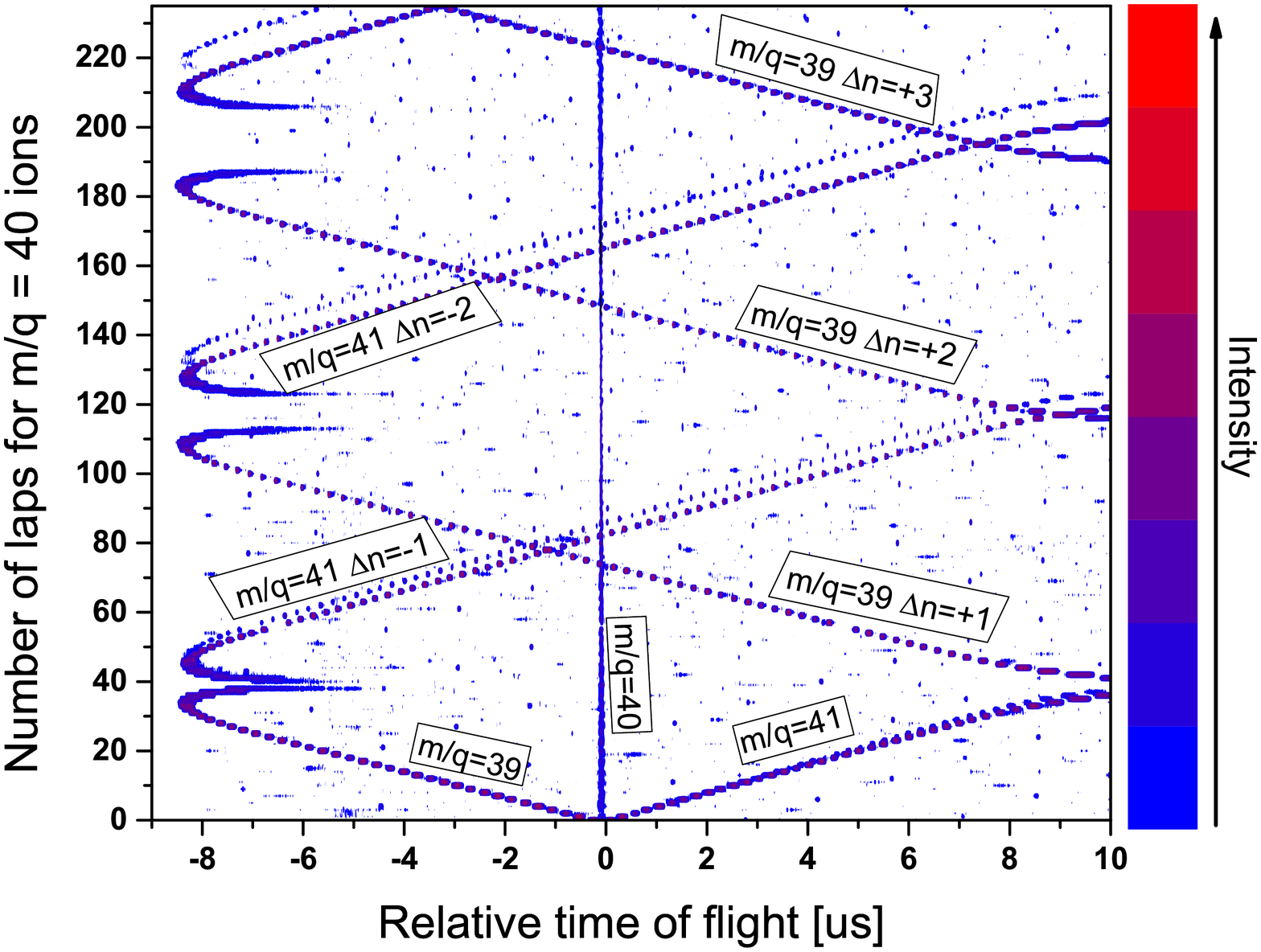}
 %}
\caption{(color online) n-vs-TOF plot centered on $m/q$=40 ions demonstrating ions of various masses making unequal numbers of laps.  Each spectrum in the plot represents 10 minutes of accumulated spectra.
\newline
\indent{¡¡¡¡} It is worth noting that the molecular isobars with $m/q$=41 become separated by $n$=50 laps.  The $m/q$=39 and 40 ions are $^{39}$K$^+$ and NaNH$_3^+$, respectively, while the $m/q$=41 ions are $^{41}$K$^+$ and NaH$_2$O$^+$.  The ``noise" peaks in the n-vs-ToF plot are caused by ions with $m/q$ very different from the central value.}
\label{fig-mass_bandwidth}
\end{figure}

 The time at which the ejection-side mirror is opened is chosen by using
\begin{equation}
t_{\mathrm{ejec}t}=t^{(0)}_\mathrm{eject}+nB
\label{eqEjectionTime}
\end{equation}
where $B$ is the circulation time of the central species of interest and $t^{(0)}_\mathrm{eject}$ is chosen to ensure the ions are not too close to the ejection mirror at the time of ejection, as such would lead them to sample a changing electric field. For a chosen set of $t^{(0)}_\mathrm{eject}$ and $B$, one can produce a set of spectra for some range of $n$ and produce a 2D color relief plot of counts against lap number $n$ and ToF, a so-called "n-vs-ToF plot", as shown in Fig.~\ref{fig-mass_bandwidth}.

\par From such a figure, we can easily see that after not so many laps ions with mass differing by a couple percent start to make different numbers of laps.  We can also see, at the edges of the plot, the deleterious effect of the switching ejection electrode on nearby ions.  These effects will be discussed in detail in Section 3.

\par The mass of an unknown species can be determined from a single species of reference ion that has traveled a flight path identical to the unknown ion species as
\begin{equation}
m = m_{\mathrm{ref}}\left(\frac{t - t_0}{t_{\mathrm{ref}}-t_0} \right)^2,
\label{eqMassWithSingleRef}
\end{equation}
where $t$ and $t_\mathrm{ref}$ are the times of flight of the unknown species and the reference, respectively, and $t_0$ is the delay between TDC start signal and actual extraction of ions from the trap.  This method has been verified to achieve relative mass precision of $\frac{\delta m}{m}$$\leq$5$\times$10$^{-7}$ \cite{ItoLi8} with only a few hundred ions.  However, Eq.~(\ref{eqMassWithSingleRef}) requires that the reference and unknown species have both made the same number of laps in the reflection chamber.  As can be seen from Fig.~\ref{fig-mass_bandwidth}, due to a limited mass bandwidth it is often the case that peaks in a given spectrum are associated with ions that made $n + \Delta n$ laps while the reference species made $n$ laps.  A means of accurately analyzing such peaks in a spectrum would be very useful.

\section{Mass Bandwidth}
\label{secBandwidth}

\par Ions moving in the MRTOF-MS can be described in terms of runners on a circular track, whose speeds are determined by their $A$/$q$ ratio.  After some number of laps, the faster (lighter) ones will overtake and ``lap" the slower (heavier) ones.  The fraction of $m/q$ that is making the same number of laps is what we call the theoretical mass bandwidth of the MRTOF-MS.

\par Once all ions in a spectrum are no longer making identical numbers of laps in the MRTOF-MS, analysis can become difficult.  The difficulty largely arises from ambiguities being introduced due to the differing lap numbers, resulting in the peaks no longer being ordered.  For instance, consider analysis of natural Potassium.  As long as the mass bandwidth of the MRTOF-MS is wider than 5\%, the MRTOF-MS can be operated such that the peaks corresponding to $^{39,40,41}$K$^+$ will be arranged with a monotonic relationship to mass.  However, if the MRTOF-MS is operated with any smaller mass bandwidth, \emph{e.g.} 4\%, the device cannot be operated such that ordering of the peaks will be arranged with a monotonic relationship to mass.  Specifically, the order of the peaks will be some permutation of \{39, 41, 40\} (see Fig. \ref{fig-mass_bandwidth}).  

\par Because of this reordering, without sufficiently limiting the mass band sent into the MRTOF-MS, one cannot presume that the ions corresponding to peaks with a larger time-of-flight have a larger mass (or more precisely $m/q$) than those corresponding to shorter times-of-flight.  This prevents use of analysis by mass differences \cite{MassDifferenceAnalysis}\cite{Wolf} as well as other straight forward analysis techniques used to identify unknown peaks.

\par At the same time, it is generally ill-advised to overly limit the mass bandwidth prior to the MRTOF-MS.  Doing so reduces the sensitivity and  
increases the amount of analyte required. Thus, a discussion of the mass bandwidth is a useful endeavor.

\subsection{Theory}

\par The time required to travel from the trap to the detector and make $n$ laps in the MRTOF in the interim can be written as
\begin{equation}
t^{(n)} = t^{(0)}+bn \sqrt{m} = (a + bn)\sqrt{m} = (\zeta +n)b\sqrt{m}
\label{eq_tof}
\end{equation}
where $t^{(0)}$ is the time required to travel from trap to detector without any reflections, $m$ is the ion mass (or mass-to-charge ratio), $b$ is a constant given by $\oint\frac{\mathrm{d}l}{\sqrt{2K}}$, corresponding to the circulation time of a unit mass and $a$ is a constant similar to $b$ and given by $\int^{detector}_{trap}\frac{\mathrm{d}l}{\sqrt{2K}}$ for ions that make no reflections, and the ratio $\zeta=a/b$ is almost constant for any particular given voltage configuration of the MRTOF-MS. $K$ is the kinetic energy of the ion, determined by the electric fields of the reflection chamber, while d$l$ is a differential length of the flight path.  As the flight path can vary slightly from one lap to the next, $b$ (and consequently $\zeta$) is only constant on average.  The slight variability in $b$ will lead to a limit in the accuracy of wide-band mass measurements. 

\par Two species of ions with masses $m\neq m'$ may have the same time of flight if they differ in the number of laps they make by $\Delta n$,
\begin{equation}
(\zeta + n)b\sqrt{m} = [\zeta + (n+\Delta n)]b\sqrt{m'}.
\label{eq_lap_condition}
\end{equation}

From Eq.~(\ref{eq_lap_condition}), the number of laps, $n_m$, at which an ion with $m/q$=$m$ will have the same time of flight  as an ion with $m/q$=$m'$ that makes $n+\Delta n$ laps is then given by
\begin{eqnarray}
%n= \frac{a\sqrt{m_1} -a\sqrt{m_2} - \Delta nb\sqrt{m_2}}{b\sqrt{m_2} - b\sqrt{m_1}}\\
n_m=\Delta n\frac{\sqrt{m'}}{\sqrt{m'}-\sqrt{m}} - \zeta.
\label{eq_max_lap}
\end{eqnarray}

\par We now define the mass bandwidth of the MRTOF-MS as the maximum fractional mass range capable of simultaneously making the same number of reflections ($|\Delta n|$$<$1). If we rewrite $m'$ as  $m$+$m_\Delta$ and solve Eq.~(\ref{eq_max_lap}) for $m_\Delta$, in the limit $\Delta n$$\rightarrow$1, we find that the mass bandwidth can be written as
\begin{equation}
\frac{m_\Delta}{m} = \frac{2(n_m+\zeta)+1}{(n_m+\zeta)^2}.
\label{eq_mass_bandwidth}
\end{equation}

\subsection{Effective Mass Bandwidth}

\par Our MRTOF-MS is typically operated with $\zeta$$\approx$0.7.  Using Eq.~(\ref{eq_mass_bandwidth}) to determine the mass bandwidth, it may be interesting to first note that at $n$=0, the mass bandwidth is $\frac{m_\Delta}{m}$$\approx$500\%, indicating that if an ion with mass $m$ reaches the ejection-side turning point just as the mirror is opened, an ion of mass $m/6$ will have already made one lap in the MRTOF.  Our time focus is typically near $n$=125 laps, giving a mass bandwidth of $\frac{m_\Delta}{m}$$\approx$1.6\%.

\begin{figure}[h]
%\resizebox{0.54\textwidth}{!}{%
 \includegraphics[width=0.48\textwidth]{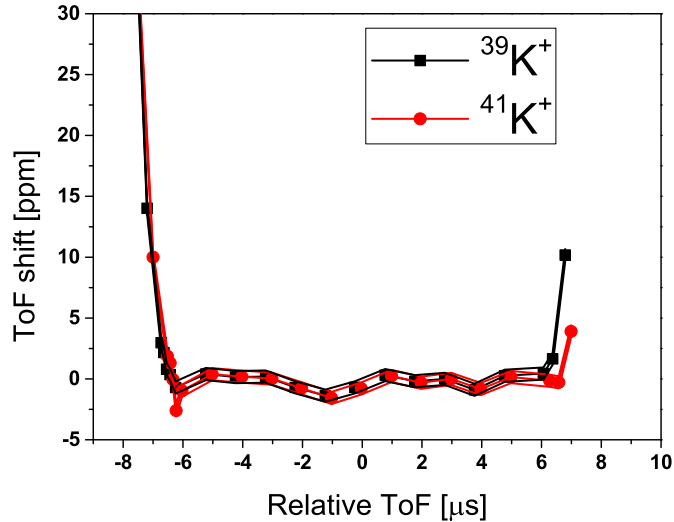}
 %}
\caption{(color online) Fractional shift in absolute time of flight as a function of the relative time of flight for $^{39,41}$K$^+$.  The longer circulation time of $^{41}$K$^+$ as compared to $^{39}$K$^+$ accounts for the slightly larger flat region for the $^{41}$K$^+$ curve.  The flat range implies that the effective mass bandwidth is $\approx$60\% of the theoretical mass bandwidth in this case. }
\label{fig-effective_mass_bandwidth}
\end{figure}

\par Unfortunately, the theoretical mass bandwidth represents an upper limit.  Implicit in our derivation was the notion that the position of ions within the MRTOF at the moment of ejection is irrelevant.  In reality, ions located sufficiently near the outermost electrode of the ejection mirror will experience a ToF shift relative to far away ions.  This comes about because ions near the electrode will experience a sharp change in the electric field at the moment of switching, while distant ions will not experience such a shift in their local electric field.  The effect is clearly visible in Fig.~\ref{fig-mass_bandwidth} for ions with a relative time of flight beyond $\pm$8~$\mu$s.

\par This effect turns out to significantly reduce the effective mass bandwidth.  Under ideal circumstances, where the switched ejection optics had no effect on the ions, we would expect that the ions would have the same absolute time of flight without regard to their relative time of flight.  In such a case, the allowable range of the relative time of flight would be $\pm$0.5$b\sqrt{m}$.

\par By systematically changing $t^{(0)}_\mathrm{eject}$ (see Eq.~(\ref{eqEjectionTime})), we could measure the ions' absolute times of flight as functions of their relative time of flight, where zero relative time of flight indicates ions in the injection-side turning point at the moment the ejection-side mirror is opened.  The result of such a measurement performed with $^{39,41}$K$^+$ ions is shown in Fig.~\ref{fig-effective_mass_bandwidth}.  We clearly see that the acceptable relative times of flight span a range of $\approx$$\pm$6~$\mu$s.   Since $\pm$0.5$b\sqrt{m}\approx$10~$\mu$s, we infer from Fig.~\ref{fig-effective_mass_bandwidth} that the effective mass bandwidth is $\approx$60\% of the theoretical mass bandwidth.

\subsection{Experimental Results}
\par We have made use of a thermal ion source to test the MRTOF-MS.  The ion source primarily produced Na$^+$ ions, as well as K$^+$ ions.  The preparation ion trap itself has a limited mass bandwidth, such that Na$^+$ ions could not be stored alongside K$^+$ ions, although NaX$^+$ molecular ions could.  The data shown in Fig.~\ref{fig-mass_bandwidth} was obtained by setting the trap to store K$^+$ ions and measuring spectra of these ions for $m/q$=40 (the center of the Potassium isotope set) ions making laps $n_{40}$$\in$[0,230].

\par Because Eqs.~(\ref{eq_max_lap}) \& (\ref{eq_mass_bandwidth}) are not intuitive, we have experimentally verified their validly.  Doing so first required precise determination of $\zeta$.  To do this, a set of high-statistics measurements of $^{39}$K$^+$ were made.  From these measurements it was determined that $a$=2\,191.345(56)~ns/$\sqrt{\mathrm{u}}$ and $b$=3\,190.229(42)~ns/$\sqrt{\mathrm{u}}$, giving $\zeta$=0.686\,893(20).

%$^{39}$K$^+$ $t^{(0)}$=13\,678.48(35)~ns while $b\sqrt{m(^{39}\mathrm{K}^+)}$=19\,913.50(26)~ns

\par Using the single-reference method described in \cite{ItoLi8}, and assuming that the $m/q$=39 ions were $^{39}$K$^+$, we determined that the $m/q$=41 ions were $^{41}$K$^+$ and NaH$_2$O$^+$ while the $m/q$=40 ions were NaNH$_3^+$.  In principle, there must also be some $^{40}$K$^+$, but, being 0.0116\% of natural Potassium, the rate is negligibly small and increasing the beam intensity so as to make it visible would result in the other ions saturating our detector.  Knowing the identities of these ions, we then used Eq.~(\ref{eq_max_lap}) to calculate the number of laps at which any pair would cross -- the number of laps after which the two ion species would begin to differ from each other by $\Delta n$ laps -- and have listed them in Table~\ref{tabLaps}.

\par
\begin{figure}[h]
%\resizebox{0.54\textwidth}{!}{%
 \includegraphics[width=0.49\textwidth]{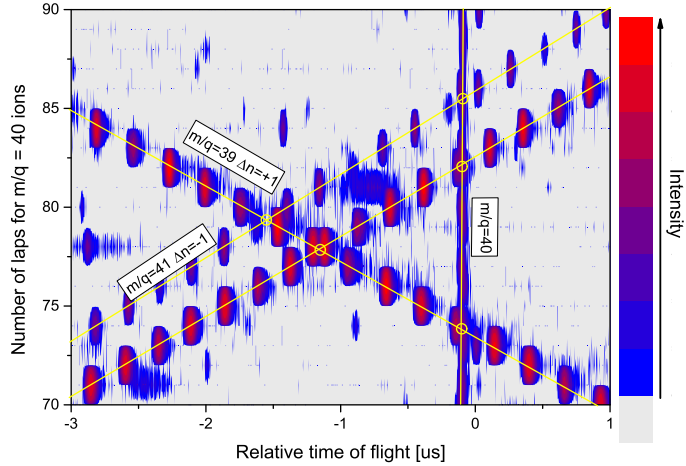}
 %}
\caption{(color online) Figure~\ref{fig-mass_bandwidth} magnified and centered on the first region wherein $m/q$=\{39,41\} cross paths with $m/q$=40.  Lines used to determine the crossing position with a precision of some fraction of a lap have been added.}
\label{fig-mass_bandwidth_lines}
\end{figure}

 To test these results, we examined Fig.~\ref{fig-mass_bandwidth} to determine the laps at which the various ion species crossed paths.  As the ion species need not perfectly overlap when they cross paths, in order to make a useful test of Eq.~(\ref{eq_max_lap}), the values for $n_\mathrm{meas}$ in Table~\ref{tabLaps} needed to be determined with a precision much better than one lap.  To do so, the dataset plotted as an n-vs-ToF plot in Fig.~\ref{fig-mass_bandwidth} was used as shown in Fig.~\ref{fig-mass_bandwidth_lines}.  By finding the centers of several peaks in the vicinity of the intersection, the intersection position can be determined to within a reasonably small fraction ($\ll$1) of one lap from the intersection of line segments.  The measurements shown in Table \ref{tabLaps} are in agreement with Eq.~(\ref{eq_max_lap}).  Let us here note that this exercise serves no purpose beyond verification of Eq.~(\ref{eq_max_lap}).

\par It is instructive to note that the calculated $n$ at which the ion species cross paths is the value of $n$ corresponding to ion species m$_1$, while the measured $n$ obtained from Figs.~\ref{fig-mass_bandwidth}~\&~\ref{fig-mass_bandwidth_lines} correspond to NaNH$_3^+$.  As an example, $^{39}$K$^+$ and $^{41}$K$^+$ were found to cross with $\Delta n$=-2 at $n_{40}$=79.3.  However, Eq.~(\ref{eq_max_lap}) yields $n_{39}$=80.3.  Examining Fig.~\ref{fig-mass_bandwidth} we see that at this crossing point, $^{39}$K$^+$ makes one lap more than NaNH$_3^+$, thus $n_{39}$=80.3.

\begin{table}[htdp]
\caption{Laps at which ions within $m/q$=\{39, 40, 41\} cross for $\Delta n$ = $n_1 - n_2$.  For rows with m$_1$ being $^{39}$K$^+$, the values of $n_\mathrm{meas}$ must be adjusted to compensate for $^{39}$K$^+$ making fewer laps than NaNH$_3^+$.  See Fig.~\ref{fig-mass_bandwidth} and the text for details.}
\begin{center}
\begin{tabular}{|c|c|c|l|l|}
\hline
m$_1$ & m$_2$ & $\Delta n$ &  $n_\mathrm{calc}$(m$_1$) & $n_\mathrm{meas}$(NaNH$_3^+$) \\
\hline
$^{39}$K$^+$ & NaH$_2$O$^+$ & -2 & ~~78.83 & ~~77.9 + 1 \\
$^{39}$K$^+$ & NaH$_2$O$^+$ & -4 & 158.34 & 156.4 + 2 \\
$^{39}$K$^+$ & NaH$_2$O$^+$ & -5 & 198.09 & 195.3 + 3 \\

$^{39}$K$^+$ & $^{41}$K$^+$ & -2 & ~~80.30 & ~~79.3 + 1 \\
$^{39}$K$^+$ & $^{41}$K$^+$ & -4 & 161.29 & 159.2 + 2 \\
$^{39}$K$^+$ & $^{41}$K$^+$ & -5 & 201.78 & 198.9 + 3 \\

NaNH$_3^+$ & $^{39}$K$^+$ & 1 & ~~73.84 & ~~73.9 \\
NaNH$_3^+$ & $^{39}$K$^+$ & 2 & 148.37 & 148.4 \\
NaNH$_3^+$ & $^{39}$K$^+$ & 3 & 222.90 & 222.9 \\

NaNH$_3^+$ & $^{41}$K$^+$ & -1 & ~~85.45 & ~~85.5 \\
NaNH$_3^+$ & $^{41}$K$^+$ & -2 & 171.60 & 171.7 \\

NaNH$_3^+$ & NaH$_2$O$^+$ & -1 & ~~82.14 & ~~82.2 \\
NaNH$_3^+$ & NaH$_2$O$^+$ & -2 & 164.97 & 165.0 \\

\hline

\end{tabular}
\end{center}
\label{tabLaps}
\end{table}%

\par As we can see, Eq.~(\ref{eq_max_lap}) does a rather good job of determining the number of laps at which a lighter mass ion will overtake a heavier one.  In principle, one could solve Eq.~(\ref{eq_max_lap})  for $m_2$ and determine the mass of $m_2$ from a reference mass $m_1$ by precisely determining at what number of laps the lighter mass ion will overtake a heavier one.  We will now show that there is a way to utilize Eq~\ref{eq_max_lap} to analyze wide-band mass spectra.

\section{Wide-band Mass Measurements}
\label{secMass}

\par As we have just shown, at its time focus our MRTOF-MS has an effective mass bandwidth of $\approx$1\%.  While the resolving power that can be achieved at the time focus is rather high, $R_m>$10$^5$, and the device has been demonstrated to be accurately precise to the level of $\delta m$/$m<$5$\times$10$^{-7}$ with a few hundred detected ions \cite{ItoLi8}, the small mass bandwidth is a real drawback.  As can be seen in Figs.~\ref{fig-mass_bandwidth} \& \ref{fig-mass_bandwidth_lines}, ions outside the mass bandwidth can still be in the spectrum, however.  Thus it would be beneficial to be able to make some analysis of such peaks.

\par Solving Eq.~(\ref{eq_max_lap}) for $m_2$ we find
\begin{equation}
m_2 = m^{(n)}_1\left(\frac{\zeta+n_{m_1}}{\zeta+n_{m_1}+\Delta n}\right)^2
\label{eqMassWithLaps}
\end{equation}
where ions with mass $m^{(n)}_1$ make $n_{m_1}$ laps and ions with mass $m_2$ make $n_{m_1}+\Delta n$ laps.
%where $n_{m_1}$ again corresponds to the number of laps made by ions with mass $m^{(n)}_1$ and $n_{m_1}+\Delta n$ corresponds to the laps made by ions with mass $m_2$.

\par Functionally, Eq.~(\ref{eqMassWithLaps}) requires that $m^{(n)}_1$ and $m_2$ have the exact same time of flight.  If we have one reference species in the spectrum, for which we know the identity and number of laps $n$, we can then calculate the mass of an artificial reference which, after $n$ laps, would have a time of flight exactly equal to that measured for the unknown species.  This artificial reference will then play the role of $m^{(n)}_1$ in Eq.~(\ref{eqMassWithLaps}).

\par
\begin{figure}[H]
 \includegraphics[width=0.48\textwidth]{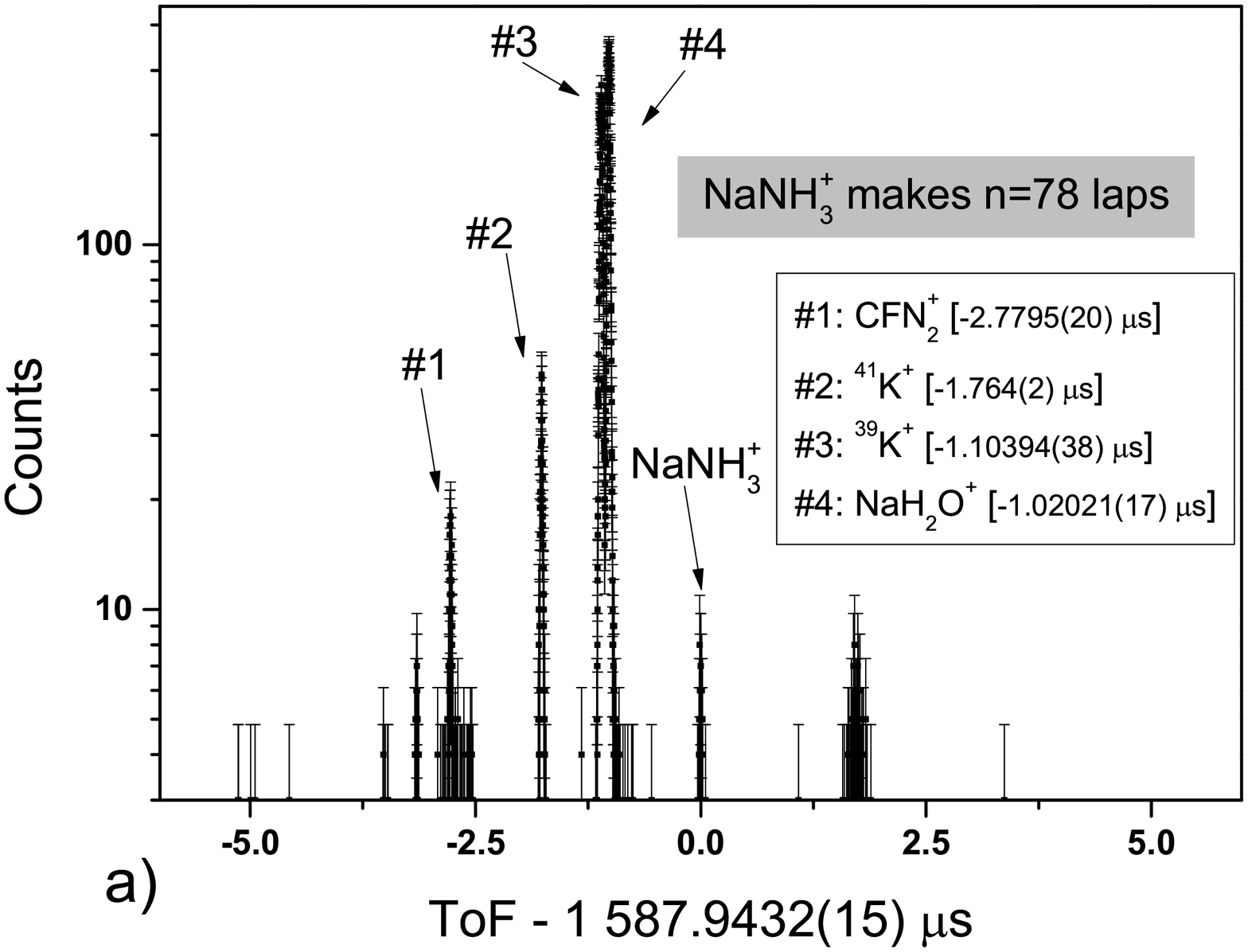}
 \includegraphics[width=0.48\textwidth]{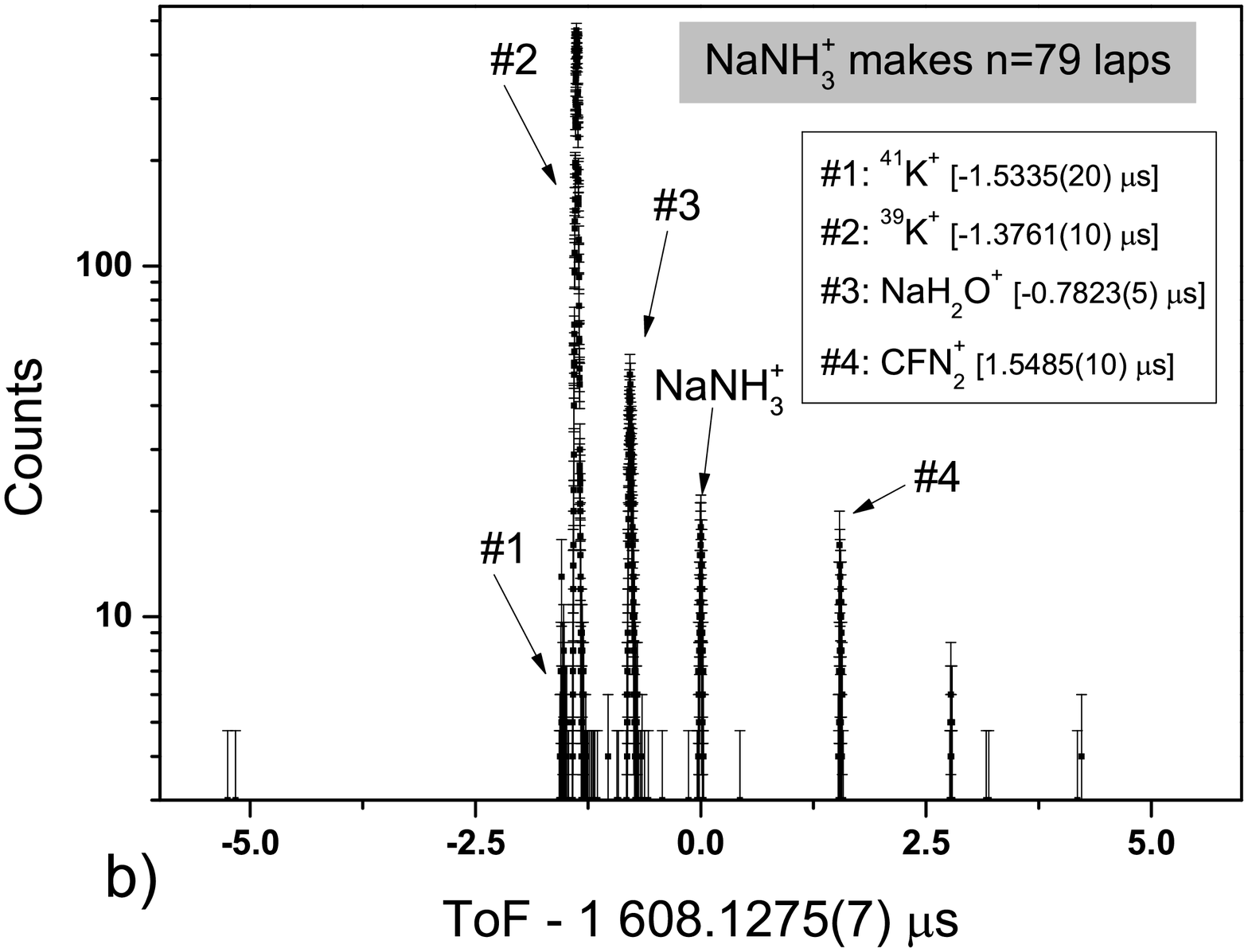}
 \includegraphics[width=0.48\textwidth]{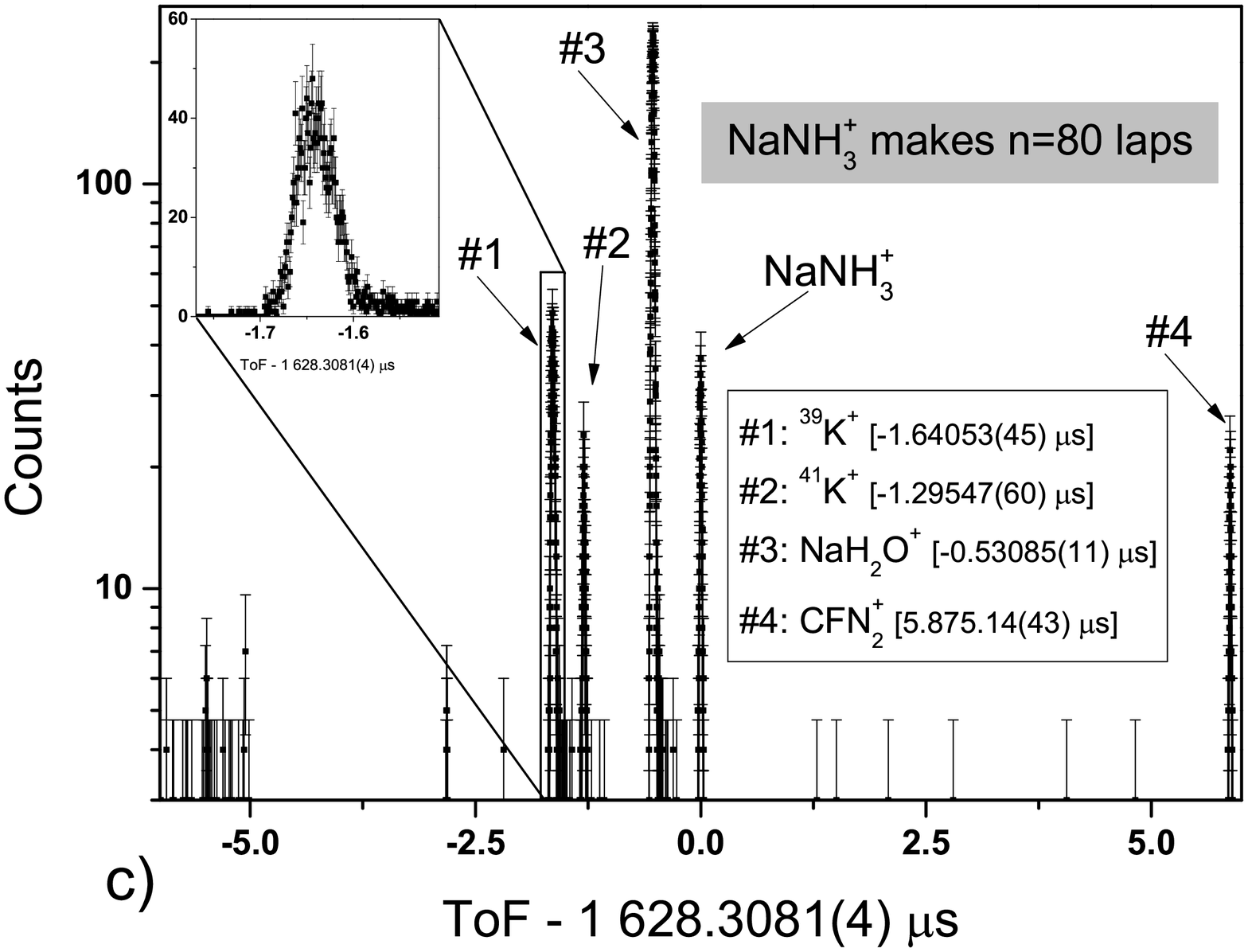}
\caption{Example spectra with $n_{40}$=78, 79, and 80 laps.  In the $n_{40}$=80 spectrum, an expansion of peak \#1 provides peak shape detail; the peak is wider than optimal as the time focus is at $\Delta n$=+35 laps.  The range is limited to the ``safe" range as determined by Fig.~\ref{fig-effective_mass_bandwidth}. The x-axis label reflects the time of flight of the NaNH$_3^+$ reference; times of flight for the peaks are relative to NaNH$_3^+$. }
\label{fig78laps}
\end{figure}

 Let us here give an example of how to use Eq.~(\ref{eqMassWithLaps}) to verify the identity of one peak using a known reference.  In Fig.~\ref{fig78laps}c, which is the $n$=80 row of Fig.~\ref{fig-mass_bandwidth}, we know that NaNH$_3^+$ has a time of flight of $t$=1\,628\,308.07(39)~ns after making 80~laps ($n_{40}$=80).  The peak labeled ``\#3" has a time of flight of $t$=1\,627\,777.15(11)~ns, but the number of laps is unknown.  If we use the single-reference method of Eq.~(\ref{eqMassWithSingleRef}) we calculate a mass of $m$=39.989\,676(10)~u.  The closest match is C$_2$O$^+$, but since that differs by 177(1)~ppm we can exclude it and conclude this peak belongs to an ion that doesn't make 80 laps.  Now we can use Eq.~(\ref{eqMassWithLaps}) with the artificial reference $m^{(80)}_1$=39.989\,676(10)~u.  If we assume $\Delta n$=-1, we find $m_2$=40.999\,644(31)~u; the closest match is NaH$_2$O$^+$ which differs in mass by 3.6(8)~ppm. The 4.5-$\sigma$ error is an effect of the fact that $b\sqrt{m}$ can vary by several nanoseconds from one lap to the next; in this case, a shift of 6~ns (a 0.03\% change in $b\sqrt{m}$) can account for the discrepancy.  In most cases, however, this would not result in ambiguity of identification.

%\par By way of example, in the data set used for Figs.~\ref{fig-mass_bandwidth} \& \ref{fig-mass_bandwidth_lines} we know that at $n_{40}$=80 laps, NaNH$_3^+$ has a time of flight of $t$=1\,628\,308.07(39)~ns.  In the same spectrum, there are four other strong peaks (see Fig.~\ref{fig78laps}).  One has a time of flight $t$=1\,627\,777.15(11)~ns.  Using the single-reference method with NaNH$_3^+$ as the reference, we calculate $m^{(80)}_1$=39.989\,676(10)~u.  Now, if we use this value of $m^{(80)}_1$ in Eq.~(\ref{eqMassWithLaps}) and assume $\Delta n$=-1, we find $m_2$=40.999\,644(31)~u, which differs from NaH$_2$O$^+$ by -141(31)~$\mu$u.  The 4.5-$\sigma$ error is an effect of the fact that $b\sqrt{m}$ can vary by several nanoseconds from one lap to the next; in this case, a shift of 6~ns (a 0.03\% change in $b\sqrt{m}$) can account for the discrepancy.
\par In general, one is presented with spectra with several unknown peaks, without any \emph{a priori} knowledge of how many laps any ions save the reference have made.  If a set of such spectra can be made wherein the reference makes a different number of laps in each spectrum, we can use Eq.~(\ref{eqMassWithLaps}) to determine candidate masses of the ions corresponding to each peak for some set of $\Delta n$.

%\par Of course, there is limited value in simply verifying the identity of an \emph{a priori} known species.  In general, one is presented with a spectrum with several unknown peaks, without any \emph{a priori} knowledge of how many laps any ion save the reference have made.  If a set of such spectra can be made wherein the reference makes a different number of laps in each spectrum, we can use Eq.~(\ref{eqMassWithLaps}) to determine candidate masses of the ions corresponding to each peak for some set of $\Delta n$.

\par One could make a table of such candidate masses for some range of $\Delta n$ and then search the table for matches within some set of spectra.  Alternatively, one can follow a straightforward algorithm to determine the mass of an ion if it has the same value of $\Delta n$ for at least two values of $n_m$.  From Eq.~(\ref{eq_max_lap}) it is readily obvious that this condition places a limit of $\Delta_m$ = 3$m$.

%\par One could make a table of such candidate masses for some range of $\Delta n$ and then search the table for matches within some set of spectra.  However, such tedious effort is not necessary.  Instead, one can follow a fairly straightforward algorithm to determine the mass of an ion if it has the same value of $\Delta n$ for at least two values of $n_m$.  From Eq.~(\ref{eq_max_lap}) it is readily obvious that this condition places a limit of $\Delta_m$ = 3$m$.

\par Consider a pair of spectra wherein the same species of reference ion makes a different number of laps, $n_m$ and $n_m'$, in each.  Initially, it is not possible to tell which of the $i$ peaks in spectrum $n_m$ correspond to which of the $j$ peaks in spectrum $n_m'$.  However, it is possible to determine the would be value of $\Delta n$ if peak $i$ did correspond to peak $j$, and in the process to determine the would be mass of peak $i$ (and peak $j$).  Doing so requires using Eq.~(\ref{eqMassWithLaps}) to calculate the masses of peak $i$ in spectrum $n_m$ and peak $j$ in spectrum $n_m'$ for some set of $\Delta n$.  Comparing them, one will find that for a particular value of $\Delta n$ the difference in the mass of peak $i$ and peak $j$ will be minimized.  Extending this to all combinations of $i$ and $j$, we can use Eq.~(\ref{eqMassWithLaps}) to build the minimized mass difference matrix

\begin{equation}
(\Delta m)_{ij} = \left|m_i^{(n)}\left(\frac{\zeta+n_m}{\zeta+n_m+\Delta n_{ij}}\right)^2 - m_j^{(n')}\left(\frac{\zeta+n_m'}{\zeta+n_m'+\Delta n_{ij}}\right)^2\right|,
\label{eqFindDn}
\end{equation}
where $m_i^{(n)}$ and $m_i^{(n')}$ are the artificial reference masses for peaks $i$ and $j$ in the spectra where the reference makes $n_{m}$ and $n_{m}'$ laps, respectively, and $\Delta n_{ij}$ is the value of $\Delta n$ that minimizes the mass difference of peaks $i$ and $j$.

\par Each row of $(\Delta m)_{ij}$ is the difference in calculated mass of peak $i$ in spectrum $n$ and peak $j$ in spectrum $n'$.  If peaks $i$ and $j$ represent the same ion species, the $j^\mathrm{th}$ column of the $i^\mathrm{th}$ row should be the minimum value in the $j^\mathrm{th}$ column and in the $i^\mathrm{th}$ row. Each row (column) of $\Delta_{ij}$ should contain no more than one column (row) minimum, as peak $i$ cannot correspond to both peak $j$ and peak $j'$.

\par Let us test this method with the set of data in the n-vs-ToF plot shown in Figs.~\ref{fig-mass_bandwidth} \& \ref{fig-mass_bandwidth_lines}.  The ions with $m/q$=\{39, 40, 41\} come near each other for the first time around $n_{40}$=80 laps.  The spectra with $n_{40}$=\{78,~79,~80\}~laps are shown in Fig.~\ref{fig78laps}.

\par After fitting the peaks in each spectra, we can use Eq.~(\ref{eqFindDn}) to identify each peak.  In principle the technique could be used for all visible peaks, however we will limit ourselves to the four most intense peaks for the sake of brevity.  

\par
\begin{figure}[H]
 \includegraphics[width=0.48\textwidth]{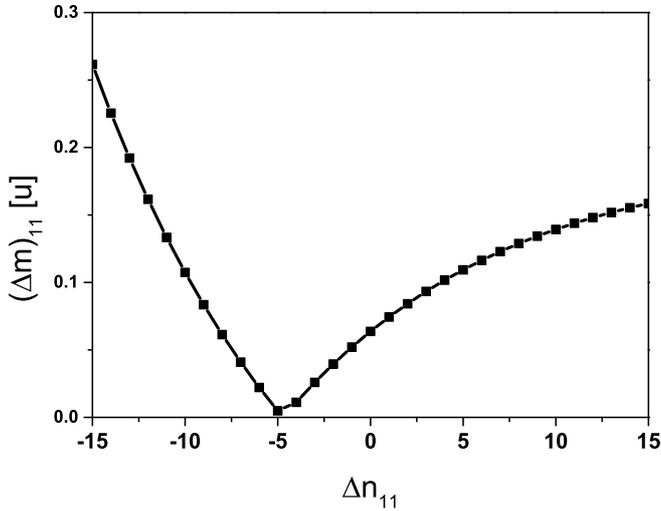}
 \vspace{-8mm}
\caption{The difference in calculated mass as a function of $\Delta n$ for peaks \#1 in the $n_{40}$=78 and $n_{40}$=79 spectra.  Assuming these two peaks correspond to the same ion species, they each differ from NaNH$_3^+$ by $\Delta n$=-5 laps.}
\label{figFindDn}
\end{figure}

 To begin, consider the $n_{40}$=\{78, 79\} spectra from Fig.~\ref{fig78laps}.  To exemplify the use of Eq.~(\ref{eqFindDn}), Fig.~\ref{figFindDn} shows $(\Delta m)_{11}$ for this pair of spectra, that is the difference in calculated mass as a function of $\Delta n$ for the first peak in each spectrum.  By performing such an analysis for each pair of peaks, we find
\[\Delta n_{ij}=
\left(
\begin{array}{cccc}
-5 & -5 & -7 & -14 \\
-1 & -2 & -4 & -11 \\
2 & 1 & -1 & -9 \\
2 & 1 & -1 & -9
\end{array}
\right)
,\]
where columns $i$ are ordered by the position of peaks in the $n_{40}$=79 spectra and rows $j$ are ordered by the position of peaks in the $n_{40}$=78 spectra.  The $(\Delta m)_{ij}$ corresponding to these $\Delta n_{ij}$ are then given by

\[~~~
\left(
\begin{small}
\begin{array}{cccc}
45.46607(4)        & 45.47498(4)        & 48.04742(4)        & \bf{59.00429(5)} \\
\bf{40.96108(3)} & 42.03062(4)        & 44.32400(4)        & 53.96265(4) \\
38.00769(3)        & \bf{38.96326(3)} & 40.99940(4)        & 50.95224(4) \\
\bf{38.00769(3)} & 38.96326(3)        & \bf{40.99940(4)} & 50.95224(4)  
\end{array}
\end{small}
\right)
\]
\[-
\left(
\begin{small}
\begin{array}{cccc}
45.47084(8)        & 45.47084(1)        & 48.04342(1)        & \bf{59.00392(1)} \\
\bf{40.96140(7)} & 42.03664(5)        & 44.31814(1)        & 53.95866(1) \\
38.00369(7)        & \bf{38.96350(5)} & 40.99550(5)        & 50.94828(1) \\
\bf{38.00771(7)} & 38.96762(5)        & \bf{40.99984(5)} & 50.95367(6) 
\end{array}
\end{small}
\right)
\]
\[\implies
\left(
\begin{small}
\begin{array}{cccc}
104.9(3.9) & 91.1(1.7) & 83.2(1.8) & \bf{6.3(1.8)} \\
\bf{7.9(3.9)} & 143.3(2.9) & 93.5(1.8) & 74.0(1.8) \\
105.2(3.8) & \bf{6.1(2.9)} & 95.0(3.0) & 77.6(1.7) \\
\bf{0.6(3.8)} & 111.9(2.9) & \bf{10.8(3.0)} & 28.2(3.0)
\end{array}
\end{small}
\right)
\mathrm{ppm}\]

\par In the matrices above, we have highlighted the values of row and column minima in $(\Delta m)_{ij}$.  We can then use the position of the minima to determine the ion masses in each spectrum.  The results of such an analysis for $n_{40}$$\in$\{78-80,153-155\} are shown in Table~\ref{tabIdentify78-80}.  The CFN$_2^+$ could not be identified in the spectra with higher number of laps.  In the above demonstration case of $n_{40}$=78 and $n_{40}$=79, the first column (and fourth row) has two values that could be minima.  However, since each row and column should have one minima, the third column requires selecting the second row of the first column as the minimum.  One could carry out the analysis allowing for the possibility that $m_{1,4}$=38.007\,71(7), but would find that it does not show up in the analysis of any other pair of spectra nor correspond to any reasonable molecule.

\par From the masses in Table~\ref{tabIdentify78-80} we were able to identify the ions $^{39,41}$K$^+$ and NaH$_2$O$^+$, which were previously identified with high-precision measurements.  Additionally, we could identify CFN$_2^+$ which had not been previously identified.  As Teflon tape is used in the ions source mounting, it is not unreasonable to find CFN$_2^+$ coming from the ion source.  From $(\Delta m)_{ij}$ we can see that it makes $\Delta n$=-14 laps compared to the NaNH$_3^+$ reference, and is likely responsible for some of the ``noise" peaks in Figs.~\ref{fig-mass_bandwidth} \& \ref{fig-mass_bandwidth_lines}.

\begin{table}[htdp]
\caption{Results of analysis of the spectra with $n_{40}$$\in$\{78-80, 153-155\}. Masses are listed in atomic mass units.  Weighted average uncertainties have been scaled by Birge ratio.  NaNH$_3^+$ is the reference species.}
\vspace{-5mm}
\begin{center}
\begin{tabular}{|c|l|l|l|l|}
\hline
$n_{40}$ & m$_1$ & m$_2$ & m$_3$ & m$_4$ \\
\hline
78  \,\,\,\,   & {\small 38.96350(8)} & {\small 40.9614(1)}  & {\small 40.99984(9)} & {\small 59.0040(2)} \\
79  \,\,\,\,   & {\small 38.96326(4)} & {\small 40.9611(1)}  & {\small 40.99940(5)} & {\small 59.0043(1)} \\
80  \,\,\,\,   & {\small 38.96339(3)} & {\small 40.96113(4)}    & {\small 40.99964(3)} & {\small 59.00446(5)} \\
153 \,\,\,\, & {\small 38.96336(5)} & {\small 40.96141(6)}    & {\small 40.99992(5)} & -- \\
154 \,\,\,\, & {\small 38.96307(2)} & {\small 40.96117(3)}    & {\small 40.99969(2)} & -- \\
155 \,\,\,\, & {\small 38.96328(2)} & {\small 40.9611(1)}  & {\small 40.99966(2)} & -- \\
\hline
%\end{tabular}
%\begin{tabular}{|c|c|c|c|c|c|}
 $<m>$ &{\small  38.96327(1)} & {\small 40.96121(2)} & {\small 40.99967(1)} & {\small 59.00434(5)}\\
 ID  & $^{39}$K$^+$ & $^{41}$K$^+$ & NaH$_2$O$^+$ & CFN$_2^+$ \\
 $m_{\mathrm{ID}}$ & {\small 38.963157} &  {\small 40.961277} &  {\small 40.999785} &  {\small 59.004002} \\
{\small $<$$\Delta m/m$$>$} & {\small 3(1) ppm} & {\small -2(1) ppm} & {\small -3(1) ppm} & {\small 6(2) ppm}\\
 %$\Delta m/m$ & {\small 2.9(1.3) ppm} & {\small -1.7(1.0) ppm} & {\small -2.8(9) ppm} & {\small 5.6(1.3) ppm}\\
\hline
\end{tabular}
\end{center}
\label{tabIdentify78-80}
\end{table}

%\begin{table}[htdp]
%\caption{Results of analysis of the spectra shown in Fig.~\ref{fig78laps} and their weighted averages.  Masses are listed in atomic mass units.}
%\begin{center}
%\begin{tabular}{|c|c|c|c|c|c|}
%\hline
%n$_1$ & n$_2$ & m$_1$ & m$_2$ & m$_3$ & m$_4$ \\
%\hline
%78\, & 79 & {\small 38.96298(3)} & {\small 40.96118(3)} & {\small 40.99960(3)} & {\small 59.00398(1)}\\
%79 & 80   & {\small 38.96336(1)} & {\small 40.96111(2)} & {\small 40.99954(2)} & {\small 59.00438(4)}\\
%78 & 80   & {\small 38.96341(1)} & {\small 40.96162(3)} & {\small 40.99967(1)} & {\small 59.00430(4)}\\
%\hline
%\end{tabular}
%\begin{tabular}{|c|c|c|c|c|c|}
% Average &{\small  38.96332(1)} & {\small 40.96129(1)} & {\small 40.99963(1)} & {\small 59.00410(1)}\\
% Formula  & $^{39}$K$^+$ & $^{41}$K$^+$ & NaH$_2$O$^+$ & CFN$_2^+$ \\
% $\Delta m$ & {\small 4.1(2) ppm} & {\small 0.3(3) ppm} & {\small -3.9(2) ppm} & {\small 1.7(2) ppm}\\
%\hline
%\end{tabular}
%\end{center}
%\label{tabIdentify78-80}
%\end{table}

\par In principle, high-precision measurements could be performed on such wide-band spectra once the value of $\Delta n$ has been determined for each peak of interest.  Depending on the set of ions identified, the high-precision measurements could be performed utilizing either one or two spectra.  In cases such as the isobaric singlets $m/q$=\{39,59\} in this set, where only a single ion species exists within the mass bandwidth, two spectra would be required -- one where the reference makes $n$ laps and one where the unknown makes $n$ laps.  For cases, such as the isobaric multiplet of $m/q$=41 in this set, where multiple ion species exist within the mass bandwidth, one member of the set can be used as the reference to determine the masses of the rest of the set with high-precision.

\par Principally, following identification of the molecular ions by means of Eq.~(\ref{eqFindDn}), one ought to make interleaved measurements of each ion species and the reference to implement the two-spectra single-reference analysis for precise identification.  However, experiment scheduling precluded such for this data set, requiring us to make use of the data of Fig.~\ref{fig-mass_bandwidth} for the entire analysis.

\par  In the data set we have been using for demonstration, two-spectra single-reference mass measurements can be made for $^{39,41}$K$^+$ and NaH$_2$O$^+$ using the previously analyzed set of spectra, as they have $\Delta n$=\{$\pm$1, $\pm$2\}.  The results of such an analysis are shown in Table~\ref{tabSingleRef}.  As can be clearly seen by comparison with Table~\ref{tabIdentify78-80}, the accuracy is much improved, owing to the elimination of deleterious effects caused by the slightly non-constant nature of $b$ with respect to $n$.

\par In the case of CFN$_2^+$, it has $\Delta n$=-14.  As each spectra in the data set required 10 minutes of data accumulation, the $n_{40}$ reference spectra were made more than 2 hours prior to each $n_{59}$ spectra containing CFN$_2^+$.   No reference measurements were made to allow an accounting of drifts caused by voltage instabilities and thermal expansion, either, precluding making any analysis of CFN$_2^+$ by the two-spectra single-reference method.

\begin{table}[htdp]
\caption{Evaluation of the data in Table~\ref{tabIdentify78-80} using the two-spectra single-reference method.  We limit the analysis to the use of spectra previously analyzed in Table~\ref{tabIdentify78-80}.  This analysis could not be performed without the peak identification and $\Delta n$ determination of the previous analysis. NaNH$_3^+$ is the reference species. } 
\label{tabSingleRef}
\begin{center}
\begin{tabular}{|c|l|l|l|l|}
\hline
$n_{40}$ & $m$($^{39}$K$^+$) & $m$($^{41}$K$^+$) & $m$(NaH$_2$O$^+$) \\
\hline
78 & 38.96321(3)   & --                       & -- \\
79 & 38.96315(3)   & 40.96139(13) & 40.99971(8) \\
80 & --                       & 40.96124(4)   & 40.99976(2) \\
153 & 38.96319(2) & --                       & -- \\
154 & --                     & --                       & -- \\
155 & --                     & 40.96132(13) & 40.99984(2) \\
\hline
$<m>$ &  38.96319(1)  & 40.96128(3)   & 40.99979(1) \\
$<$$\Delta m/m$$>$  & 0.77(31) ppm & 0.13(83) ppm  & 0.09(35) ppm \\
\hline
\end{tabular}
\end{center}
\end{table}

\par Finally, for the case of non-singular isobaric sets, each spectra can be analyzed using a single-spectra single-reference method.  In our example data, the only such case is $m/q$=41 with $^{41}$K$^+$ and NaH$_2$O$^+$.  By choosing $^{41}$K$^+$ as the reference, we can then precisely determine the mass of the member of the molecular isobar set.  The results of such an analysis are shown in Table~\ref{tab1Spec1Ref}.  Since in such an analysis all ion species involved in the analysis experienced the same environment, the scattering of the results is reduced.  Because every spectra could contribute to the analysis, the weighted uncertainty is also reduced.

\begin{table}[htdp]
\caption{Evaluation of the $m/q$=41 isobar set in Table~\ref{tabIdentify78-80} using the single-spectra single-reference method.  We limit the analysis to the use of spectra previously analyzed. $^{41}$K$^+$ is the reference species. } 
\label{tab1Spec1Ref}
\begin{center}
\begin{tabular}{|c|l|c|}
\hline
$n_{40}$ & m(NaH$_2$O$^+$) & $\Delta m/m$ [ppm] \\
\hline
  78 & 40.99972(11) & -1.7(2.7) \\
  79 & 40.99959(11) & -4.6(2.6) \\
  80 & 40.99979(3) & 0.12(75) \\
153 & 40.99979(2) & 0.08(41) \\
154 & 40.99980(2) & 0.34(39) \\
155 & 40.99980(13) & 0.4(3.2) \\
\hline
 & 40.999779(11) & -0.15(26) \\
\hline
\end{tabular}
\end{center}
\end{table}

\section{Conclusion and Outlook}

\par We have presented a derivation of the mass bandwidth of the MRTOF-MS as a function of number of laps.  The actual mass bandwidth is reduced to 60\% of the calculated value due to ions near the ejection mirror being affected by the changing electric field during ejection from the MRTOF-MS.  However, one intriguing feature of the MRTOF-MS is that species outside the mass bandwidth are not lost (although they may be adversely affected by the switching of the injection and ejection mirrors) but rather make different numbers of laps than the central reference species.

\par It is possible to take advantage of this feature to identify species outside the mass bandwidth.  By using two spectra with different values of $n_m$ for the reference ion, it was possible to identify the various species with an accuracy of $\sim$5 ppm.  By using more spectra, the accuracy could be improved.

\par It is well-known that the circulation time of ions can differ slightly from one lap to the next.  We posit that this is the source of the observed deviations.  Presumably it is the effect of ions having a trajectory that crosses the axis of the MRTOF-MS, resulting in slightly different paths from one lap to the next.  The effect could, presumably, be reduced or eliminated by improving the injection optics, to ensure that the ions enter the reflection chamber of the MRTOF-MS more perfectly on axis.

\par Even with the limited accuracy of the wide-band analysis, it is sufficient to reasonably identify the various ions and determine the number of laps they make.  With such information, it is then possible to perform precision mass determinations through a two-spectra single-reference analysis, resulting in relative precision and accuracy of better than 1~ppm.  In cases of where multiple ion species exist within the mass bandwidth, such as isobaric multiplets, single-spectra single-reference analysis can be performed, resulting in relative precision and accuracy on level of 0.1~ppm or better.  We believe this analysis method should allow MRTOF-MS to not only be of value to the nuclear physics community, but to become a useful instrument for mass spectrometry as a whole.

\par
\begin{figure}[H]
 \includegraphics[width=0.48\textwidth]{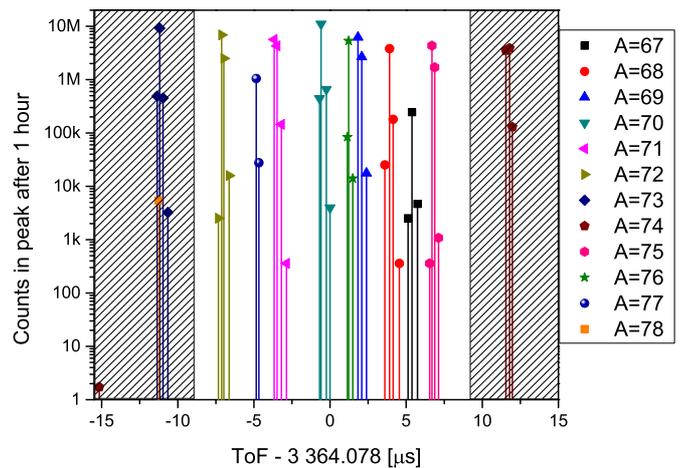}
 \vspace{-8mm}
\caption{Calculated distribution of peaks from nuclei produced by in-flight fission of Uranium, thermalized in a gas cell and analyzed by the MRTOF-MS.  Ions in the hashed area cannot be analyzed because they will be deleteriously affected by switching electrodes.}
\label{fig70Fe}
\end{figure}

 In the near future, we hope to use the MRTOF-MS to perform wide-band mass measurements of exotic radioactive nuclei.  Using in-flight fission and fragmentation of nuclei \cite{frag}, many radioactive ions can be simultaneously produced at high-energy.  After thermalizing these nuclear ions in Helium gas cell \cite{WadaGC}, they can be analyzed with the MRTOF-MS.  Using the method we have presented, we envision a highly economical use of such beams by utilizing the MRTOF-MS as a high-precision wide-band mass analyzer.  Figure \ref{fig70Fe} shows a calculated peak distribution for one such cocktail, centered on $^{70}$Fe.  Using the methods described in this manuscript we plan to identify each peak from a single well-known reference, then precisely determine each mass from one well-known reference in each isobaric set.

\par The authors wish to acknowledge the support of the Nishina Center for Accelerator Sciences.  This work was supported by the Japan Society for the
Promotion of Science KAKENHI (grants \#2200823, \#24224008 and \#24740142).



%% References


\bibliographystyle{elsarticle-num}



\end{document}